\documentstyle[12pt,a4,amssymb]{article}

	\newcommand{\be}{\begin{equation}}
	\newcommand{\ee}{\end{equation}}

	\newcommand{\bea}{\begin{eqnarray}}
	\newcommand{\eea}{\end{eqnarray}}

	\def\bra#1{\left\langle#1\,\right|}
	\def\ket#1{\left.#1\right\rangle}

\begin{document}

\begin{titlepage}

\begin{flushright}
{\tt FTUV/94-27\\
     IFIC/94-24\\
     hep-th/9406001}\\
\end{flushright}

\vfill

\begin{center}

{\bf{\Large A (1+3) Relativistic Harmonic Oscillator Simulated
	    by an Anti-de Sitter Background}}\footnote[2]{Work partially
            supported by the {\it C.I.C.Y.T.} and the {\it D.G.I.C.Y.T.}}

\bigskip
\bigskip
\bigskip

	 D.J.Navarro and J.Navarro-Salas.\\

\bigskip

\begin{center}
       Departamento de F\'{\i}sica Te\'orica and\\
       IFIC, Centro Mixto Universidad de Valencia-CSIC.\\
       Facultad de F\'{\i}sica, Universidad de Valencia,\\
       Burjassot-46100, Valencia, Spain.
\end{center}

\bigskip
\today
\bigskip

\end{center}


\begin{center}
 {\bf Abstract}
\end{center}

	It is shown that a static $(1+3)$ anti-de Sitter metric defines, in a
natural way, a relativistic harmonic oscillator in Minkowski space.
The quantum theory can be
solved exactly and leads to wave functions having a significantly different
behaviour with respect to the non-relativistic ones. The energy spectrum
coincides, up to the ground state energy, with that of the non-relativistic
oscillator. \\

\begin{flushleft}
PACS number(s): 03.65.Ge, \, 11.10.Qr, \, 04.20.-q
\end{flushleft}

\vfill

\end{titlepage}

\newpage

\section{Introduction}

\hspace{5mm} The non-relativistic oscillator is one of the simplest and
most useful system in physics. However, despite of its simplicity,
there is not a well-established relativistic generalization in the
literature. The first proposal for a relativistic harmonic oscillator
was given by Yukawa \cite{Yuk}, followed by the work of Feynman et al.
\cite{Fey} and further developed in \cite{Kim}. These works are all
based in the naive cova\-riant genera\-li\-sation $x^{\mu}x_{\mu}$ of the
non-relativistic potential thus leading to quantum timelike exitations,
the interpretation of which presents some difficulties. On the other
hand, It\^{o} et al. \cite{Ito} (see also \cite{Mos,Benit}), introduced a
Dirac equation which is li\-near in both coordenates and momenta. In the
non-relativistic limit, the equation sa\-tis\-fied by the large components
is that of ordinary oscillator whith a spin-orbit coupling term.\\

	Recently a new proposal for the relativistic harmonic
oscillator was outlined in Ref. \cite{Aldaya1} (see also \cite{Aldaya2}).
It is based in the natural generalization of the symmetry algebra of
quantum operators of a relativistic free system (i.e., the Poincar\'{e}
algebra)
\be
[\hat{E} , \hat{x}] = -i \frac{\hbar}{m} \hat{p} \, ,
\; \; [\hat{E} , \hat{p}] = 0 \, ,
\; \; [\hat{x} , \hat{p}] = i \hbar (1 + \frac{1}{m c^2}
\hat{E}) \, ,
\ee
and that of a non-relativistic harmonic oscillator (i.e., the Lie algebra
of the Newton group)
\be
\label{alg2}
[\hat{E} , \hat{x}] = -i \frac{\hbar}{m} \hat{p} \, , \; \; [\hat{E} ,
\hat{p}] = i m \omega^2 \hbar \hat{x} \, , \; \; [\hat{x} , \hat{p}] = i
\hbar \, .
\ee
$\hat{E}$, $\hat{p}$ and $\hat{x}$ are the energy (with the rest-mass energy
substracted), momentum and boost operators in the center of momentum frame.
The proposed symmetry for the $(1+1)$ relativistic
oscillator was defined in terms of the unique Lie algebra which allows to
be contracted to the above algebras
\be
\label{alg3}
[\hat{E} , \hat{x}] = -i \frac{\hbar}{m} \hat{p} \, , \; \; [\hat{E} ,
\hat{p}] = i m \omega^2 \hbar \hat{x} \, , \; \; [\hat{x} , \hat{p}] = i
\hbar (1 + \frac{1}{m c^2} \hat{E}) \, .
\ee
This Lie algebra corresponds to that of the $SO(1,2)$ group. Related
approaches can be seen in (\cite{referee}) \\

        The aim of this paper is to further elaborate on this proposal.
In Sect.2 we shall interpret it geometrically
showing that a static $(1+1)$ anti-de Sitter metric can be used to
simulate a one-dimensional relativistic oscillator.
The results obtained in this way can be seen as complementary to those
found by group theoretical methods (\cite{Aldaya3,Aldaya4,Navarro})
for the one-dimensional oscillator \cite{Aldaya1,Aldaya5}. The main goal of
this paper is to extend to three spatial dimensions the proposal (3) for a
$(1+1)$
relativistic oscillator. This will be done in Sect.3 making use
of the geometrical interpretation developed in Sect.2.
The quantum theory will be obtained by
solving the corresponding Kein-Gordon equation and leads to wave functions
having an appreciably different behaviour with respect to the non-relativistic
ones. In Sect.4 we shall state our conclusions. \\
\\
\\

\section{Anti-de Sitter space and the one-dimensional
	 relativistic oscillator}

\hspace{5mm} The basic ingredient in the proposal of Ref.\cite{Aldaya1} for
a relativistic oscillator is the $SO(1,2)$ group symmetry. The Lie
algebra commutators of this group can be throught of as the natural
generalization of those of the relativistic free particle and the
non-relativistic oscillator. Due to anti-de Sitter (AdS) space is a homogeneous
space of the $SO(1,2)$ group it is therefore natural to regard the harmonic
oscillator interaction in $(1+1)$ Minkowski space as equivalent to a free
system in AdS space (more precisely, in its universal covering space,
which has the topology of ${\Bbb R}^2$) \cite{Hawk}.
However, this definition for a
relativistic oscillator is incomplete. Owing to general covariance
we must specify which particular AdS metric properly simulates the
relativistic oscillator interaction in Minkowski space.
To solve this ambiguity we can resort to the
non-relativistic limit. To adjust the non-relativistic limit we have to
choose $g_{00}$ as
\be
g_{00} = 1 + \frac{\omega^2}{c^2} x^2 \, .
\ee
The requirement of having an AdS geometry, in particular the condition of
cons\-tant scalar curvature, determines the remaining components of the metric
\be
\label{metric}
ds^2 = (1 + \frac{\omega^2}{c^2}x^2) c^2dt^2 - \frac{1}{1+ \frac
       {\omega^2}{c^2}x^2} \, dx^2 \, .
\ee
The geodesic trajectories of motion can be obtained from the lagrangian
\be
\label{lagrangian}
{\cal L} = -mc \,\sqrt{1 - \frac{1}{1+ \frac{\omega^2}{c^2} x^2}
	   \frac{v^2}{c^2} + \frac{\omega^2}{c^2} x^2} \, ,
\ee
where $m$ is the (reduced) mass of the system. The underlying $SO(1,2)$
symmetry is realized by Poisson brackets between the three constants of motion
\be
m g_{\mu \nu} \; f^{\mu}_{(a)} \, \frac{d x^{\nu}}{d \tau} \, ,
\ee
where $f^{\mu}_{(a)}$, $a = 1,2,3$ are the Killing vectors of (\ref{metric})
\bea
f_{1}^{\mu} & = & (1,0) \, \\
f_{2}^{\mu} & = & \left( \frac{\frac{\omega}{c} x}{\sqrt{1 + \frac{\omega^2}
{c^2} x^2}} \, \cos \omega t , \sqrt{1 + \frac{\omega^2}{c^2} x^2} \, \sin
\omega t \right) \, \\
f_{3}^{\mu} & = & \left( - \frac{\frac{\omega}{c} x}{\sqrt{1 + \frac{\omega^2}
{c^2} x^2}} \, \sin \omega t , \sqrt{1 + \frac{\omega^2}{c^2} x^2} \cos
\omega t \right) \, .
\eea
The Killing vectors  realize the algebra (\ref{alg3}).
$f_{1}^{\mu}$, $f_{2}^{\mu}$ and $f_{3}^{\mu}$ lead to the energy,
boost and momentum generators respectively.
It is also interesting to note that
the periodic character of the trajectories of motion of the oscillator can be
traced back to the existence of closed time-like lines in AdS space. \\

	From (\ref{lagrangian}) it is straighforward to compute the hamiltonian
\be
\label{hamiltonian}
H^2 = m^2 c^4 + p^2 c^2 + m^2 \omega^2 c^2 x^2 + 2
      \omega^2 x^2 p^2 + \frac{\omega^4}{c^2} x^4 p^2 \, .
\ee
The quantum wave fuctions can be obtained from the Schr\"{o}dinger-type
equation associated with (\ref{hamiltonian}). With the standard
substitutions $H \rightarrow i \hbar \frac{\partial}{\partial t}$,
$x \rightarrow x$, $p~\rightarrow~-i \hbar \frac{\partial}{\partial x}$
and introducing the parametres $\alpha$, $\beta$ to account for the
normal ordening ambiguities of the classical function (\ref{hamiltonian})
\bea
x^4 p^2 \; & \longrightarrow & \; - \hbar^2 (x^4 \frac{d^2}{dx^2} + 4 x^3
     \frac{d}{dx} + \alpha x^2) \, ,\\
x^2 p^2 \; & \longrightarrow & \; - \hbar^2 (x^2 \frac{d^2}{dx^2} + 2 x
     \frac{d}{dx} + \beta) \, ,
\eea
the Schr\"{o}dinger equation implies the Klein-Gordon equation in AdS space
\cite{Bir}
\be
\label{kg}
(\square + \frac{m^2 c^2}{\hbar^2} + \xi R) \phi = 0 \, ,
\ee
where $\square$ is the D'Alembertian operator for the AdS
metric (\ref{metric}), $R = -2 \frac{\omega^2}{c^2}$ the scalar
curvature and $\xi$ a numerical factor. The equivalence is obtained
through the transformation
\be
\label{transf}
\Psi = \frac{1}{\sqrt{1+\frac{\omega^2}{c^2}x^2}} \, \phi \, ,
\ee
and requires a restriction on the parametres $\alpha$ and $\beta$:
$\alpha = \xi + \frac{1}{2}$, and $\beta = 2\xi + 2$.\\

	It is worthwhile to remark that the thansformation (\ref{transf})
can be understood in terms of the Klein-Gordon scalar product in curved
space (see, for instance \cite{Bir})
\be
\label{ps1}
\left \langle \phi_1 \, \right |  \left.  \phi_2 \right \rangle = i
\int_{\Sigma} d \sigma_{\mu} \, \sqrt{g} \, g^{\mu\nu}(\phi_1 \stackrel{
\leftrightarrow}{\partial_\nu} \phi_{2}^{\ast}) \, ,
\ee
where $\Sigma$ is the initial value hypersurface. For the line element
(\ref{metric}) and choosing $\Sigma$ as $t=0$, (\ref{ps1}) becomes
\be
\label{ps2}
\bra{\phi_1} \ket{\phi_2} = -i \int \frac{dx}{1 + \frac{\omega^2}{c^2}
x^2} (\phi_1 \stackrel{\leftrightarrow}{\partial_0} \phi_{2}^{\ast}) \, .
\ee
For stationary states the scalar product (\ref{ps2}) is proportional to
the standard scalar product of Schr\"{o}dinger wave functions
\be
\bra{\Psi_1} \ket{\Psi_2} = \int dx \, \Psi_{1} \Psi_{2}^{\ast} \, .
\ee
{}From now on we shall be mainly concerned with the Klein-Gordon
equation to study the quantum theory of the relativistic oscillator. We
shall also keep free the parameter $\xi$.\\

	The D'Alembertian operator for the AdS metric
(\ref{metric}) is
\be
\square = \frac{1}{1 + \frac{\omega^2}{c^2} x^2} \frac{1}{c^2} \frac
{\partial^2}{\partial t^2} - 2\frac{\omega^2}{c^2} x \frac{\partial}
{\partial x} - (1 + \frac{\omega^2}{c^2} x^2) \frac{\partial^2}
{\partial x^2} \, .
\ee
To solve the Klein-Gordon equation we shall first look for positive
frequency states. To find the spatial dependence of the wave
functions we shall further propose the ansatz
\be
\label{sol}
\phi (x,t) = e^{-i \lambda \omega t} \; (1 + \frac{\omega^2}{c^2}
x^2)^{- \frac{\lambda}{2}} \; \phi^{\lambda}(x) \, ,
\ee
where $\lambda$ is an arbitrary positive parameter. The wave function
(\ref{sol}) verifies the Klein-Gordon equation (\ref{kg}) if the function
$\phi^{\lambda}(x)$ satisfies the following differential equation
\be
\label{edif}
\left\{ (1 + \frac{\omega^2}{c^2} x^2) \frac{d^2}{d x^2} - 2 \frac
{\omega^2}{c^2} (\lambda - 1) x \frac{d}{dx} + \frac{\omega^2}{c^2}
\left( \lambda (\lambda - 1) - N^2 + 2 \xi \right) \right\}
\phi^{\lambda}(x) = 0 \, ,
\ee
where $N = \frac{m c^2}{\hbar \omega}$. In terms of the variable
$w = - \frac{\omega^2}{c^2} x^2$, (\ref{edif}) is a
standard hypergeometric equation
\be
\label{hypergeo}
\left\{ (1 - w) w \frac{d^2}{d w^2} + \left( \frac{1}{2} - (\frac{3}{2}-
\lambda) w \right) \frac{d}{dw} - \frac{1}{4} \left( \lambda(\lambda - 1)
- N^2  + 2 \xi \right) \right\} \phi^{\lambda} (w) =0 \, .
\ee
The regularity condition at infinity restricts the allowed values of
the $\lambda$ parameter. We obtain
\be
\label{landa}
\lambda = \frac{1}{2} + n + \gamma \, ,
\ee
where $n = 0, 1, 2, \ldots$ \, \, and
\be
\label{gamma}
\gamma = \frac{1}{2} \sqrt{1 + 4 N^2 - 8 \xi} \, .
\ee
The energy spectrum is then
\be
\label{energyone}
E = \left( \frac{1}{2} + n + \frac{1}{2} \sqrt{1 + 4 \frac{m^2 c^4}
{\hbar^2 \omega^2} - 8 \xi} \; \right) \, \hbar \omega \, .
\ee
We have chosen the positive sign of the square root of (\ref{gamma}) to
fit the non-relativistic aproximation of (\ref{energyone}).
The above equation makes clear the physical meaning of the $\xi$
parameter. It is related with the zero-point energy of system. \\

	Introducing the variable $z = -i \frac{\omega}{c}$ and for the
discrete values (\ref{landa}) the equation (\ref{edif})
turns out to be the equation of Gegenbauer polynomials \cite{Abram}
\be
\left\{ (1 - z^2) \frac{d^2}{dz^2} - \left( 1 - 2(n + \gamma) \right) z
\frac{d}{dz} - n (n + 2 \gamma) \right\} \phi^{\lambda}_{n} (z) = 0 \, .
\ee
Therefore the Klein-Gordon energy-eigenstates are
\be
\phi^{\gamma}_{n}(x,t) = e^{-i \frac{E}{\hbar} t} \; (1 + \frac{\omega^2}{c^2}
x^2)^{- \frac{\lambda}{2}} \; C^{-(n + \gamma)}_{n}(-i \frac{\omega}
{c} x) \, ,
\ee
and, in terms of the Schr\"{o}dinger wave functions (see(\ref{transf})),
they are
\be
\Psi^{\gamma}_{n}(x,t) = N _{n}^{\gamma} \; e^{-i \frac{E}{\hbar} t} \;
(1 + \frac {\omega^2}{c^2} x^2)^{- \frac{\lambda + 1}{2}} \; C^{-(n +
\gamma)}_{n}(-i \frac{\omega}{c} x) \, ,
\ee
where $N_{n}^{\gamma}$ are normalization constants
\be
\left( N_{n}^{\gamma} \right)^{-2} = \sqrt{\frac{\hbar \pi}{m \omega}}
\; \sqrt{N} \; \, \frac{4^n}{n!} \; \frac{(2 \gamma)!}
{(2 \gamma + n)!} \left[ \frac{(\gamma + n)!}{(\gamma)!} \right]^2 \frac{1}
{\gamma + n + \frac{1}{2}} \; \frac{\Gamma(\gamma + 1)}{\Gamma(\gamma +
\frac{1}{2})} \, .
\ee
We must stress that the Hermite polynomials $H_{n}(\zeta)$, where $\zeta =
\sqrt{\frac{m \omega}{\hbar}} x$, are naturally recovered in the
non-relativistic limit
\be
\label{relativistic hermite}
\lim_{c \rightarrow \infty} \frac{i^{n} n!}{N^{\frac{n}{2}}} C_{n}^{-(n +
\gamma)}(-i \frac{\zeta}{\sqrt{N}}) = H_{n}(\zeta) \, ,
\ee
we also have
\be
\label{limite}
\lim_{c \rightarrow \infty} \frac{N^n}{(n!)^2} \; \left( N_{n}^{\gamma}
\right)^2 = \sqrt{\frac{m \omega}{\hbar \pi}} \; \frac{1}{n! 2^n} \, ,
\ee
where the r.h.s. of (\ref{limite}) are the normalization constants of the
Hermite polynomials. Therefore, the l.h.s. of (\ref{relativistic hermite}),
without the limit,
can be seen as a relativistic generalization of the Hermite polynomials. In
fact, it is not difficult to check that they are proportional to the so-called
relativistic Hermite polynomials of Ref. (\cite{Aldaya5,Aldaya1}).\\

	To finish this section we would like to comment on the issue of
the zero-point energy. From a group theoretical point of view the quantum
wave functions should provide a carrier space for irreducible lowest weight
representations
of the symmetry group $SO(1,2)$. The dimensionless parameter $\frac{E_0}
{\hbar \omega} = \frac{1}{2} + \gamma$ is the lowest weight and
characterizes the representation. For the $(1+1)$ relativistic oscillator
we have $\frac{E_0}{\hbar \omega} > \frac{1}{2}$. This mean that not all the
lowest weight representations $(\frac{E_0}{\hbar \omega}~\in~[0 , + \infty[)$
can be realized physically. The natural barrier $\frac{E_0}{\hbar \omega} =
\frac{1}{2}$ corresponds to the so-called Mock representation \cite{Lang}. \\
\\
\\

\section{The three-dimensional relativistic oscillator}

\hspace{5mm} In this section we shall extend our study of the
one-dimensional relativistic oscillator to the three-dimensional
case. To this end we shall first find out the appropiate form of the $(1+3)$
metric. Imposing the non-relativistic limit and using the spherical symmetry
of the $(1+3)$ AdS space we can write
\be
\label{metric3}
ds^2 = (1 + \frac{\omega^2}{c^2} r^2) c^2 dt^2 - F(t,r) dr^2 -
G(t,r) \, (d \theta^2 + \sin^2 \theta d \varphi^2) \, ,
\ee
where $0 \leq \theta \leq \pi$, $0 \leq \varphi \leq 2 \pi$ are the
usual spherical coordinates. Moreover, to recover the one-dimensional
oscillator when the angular coordinates are frozen we have to choose the
function $F$ as follows
\be
F(t,r) = \frac{1}{1 + \frac{\omega^2}{c^2} r^2} \, .
\ee
Imposing now the anti-de Sitter geometry we find that the
appropriate line element should read as
\be
\label{metric3d}
ds^2 = (1 + \frac{\omega^2}{c^2} r^2) c^2 dt^2 - \frac{d r^2}{1 +
\frac{\omega^2}{c^2} r^2} - r^2 (d \theta^2 + \sin^2 \theta d
\varphi^2) \, .
\ee
The geodesics of (\ref{metric3d}) can be derived from the lagrangian
\be
{\cal L} = -mc \, \sqrt{1 + \frac{\omega^2}{c^2} r^2 - \frac{v^2}{c^2}
+ \frac{\omega^2}{c^4} \frac{(\vec{x} \cdot \vec{v})^2}{1 + \frac
{\omega^2}{c^2} r^2}} \, ,
\ee
and, according with our scheme, we can view the above lagrangian as defi\-ning
a three-dimensional relativistic oscillator in Minkowski space. The
$SO(3,2)$ symmetry  (i.e., the generalization of the
three-dimensional Poincar\'{e} and Newton algebras)
\be
[\hat{E} , \hat{x}^{i}] = -i \frac{\hbar}{m} \hat{p}^{i} \, , \; \; [\hat{E} ,
\hat{p}^{i}] = i m \omega^2 \hbar \hat{x}^{i} \, , \; \; [\hat{x}^{i} ,
\hat{p}^{j}] = i \hbar (1 + \frac{1}{m c^2} \hat{E}) \delta^{ij} \, .
\ee
(we have ommitted the rotation generators)
can be realized now by the Killing vectors of (\ref{metric3d}). \\

	The next step now is to compute the hamiltonian. We obtain
\be
\label{hamilt3}
H^2 = (1 + \frac{\omega^2}{c^2} r^2) \, (m^2 c^4 + p^2 c^2 +
      \omega^2 (\vec{x} \cdot \vec{p})^2) \, .
\ee
As in the one-dimensional oscillator it is not difficult to test
that the Schr\"{o}din\-ger equation associated with the hamiltonian
(\ref{hamilt3}) can be transformed, with a particular normal-ordering
prescription, into a Klein-Gordon equation. Introducing the
parameters $\alpha$ , $\beta$ , $\eta$ , $\varrho$ for the operator ordering
ambiguities of the classical function (\ref{hamilt3})
\bea
x_{i}^{4} p_{i}^{2} \; & \longrightarrow & \; - \hbar ^2 (x_{i}^{4} \frac
{\partial^2}
{\partial x_{i}^{2}} + 4 x_{i}^{3} \frac{\partial}{\partial x_{i}} +
\alpha x_{i}^{2}) \, , \\
x_{i}^{2} p_{i}^{2} \; & \longrightarrow & \; - \hbar^2 (x_{i}^{2} \frac
{\partial^2}
{\partial x_{i}^{2}} + 2 x_{i}^{2} \frac{\partial}{\partial x_{i}} +
\beta) \, , \\
x_{i}^{3} p_{i} \; & \longrightarrow & \; -i \hbar (x_{i}^{3} \frac{\partial}
{\partial x_{i}} + \eta) \, , \\
x_{i} p_{i} \; & \longrightarrow & \; -i \hbar (x_{i} \frac{\partial}
{\partial x_{i}} + \varrho) \, ,
\eea
the Schr\"{o}dinger equation leads to the Klein-Gordon equation in AdS
space with metric (\ref{metric3d}) ($R = - 12 \frac{\omega^2}{c^2}$). The
Schr\"{o}dinger and Klein-Gordon wave functions are related by
\be
\Psi = \frac{1}{\sqrt{1 + \frac{\omega^2}{c^2} r^2}} \, \phi \, .
\ee
The normal ordering parameters are then fixed as
\be
\alpha = \frac{11}{2} + 8 \xi \, , \; \hfill \beta = \frac{1}{2} + 2 \xi
\, , \; \hfill \eta = \frac{3}{2} \, , \; \hfill \varrho = \frac{1}{2}
\, .
\ee
So, there is only one free parameter left, which is essentially the
curvature factor $\xi$. \\

	Now we want to solve the corresponding Klein-Gordon equation. The
D'Alem\-bertian operator is given by
\be
\square = \frac{1}{1 + \frac{\omega^2}{c^2} r^2} \frac{1}{c^2}\frac
{\partial^2}{\partial t^2} - 2 \frac{\omega^2}{c^2} r \frac{\partial}
{\partial r} - (1 + \frac{\omega^2}{c^2} r^2) \frac{1}{r^2} \frac{\partial}
{\partial r} (r^2 \frac{\partial}{\partial r}) + \frac{\vec{L}^2}{r^2} \, ,
\ee
where $\vec{L}^2$ is the orbital angular momentum operator
\be
\vec{L}^2 = - \frac{1}{\sin \theta} \frac{\partial}{\partial \theta} \left(
\sin \theta \frac{\partial}{\partial \theta} \right) - \frac{1}{\sin^2
\theta} \frac{\partial^2}{\partial \varphi^2} \, . \\
\ee
Separation of variables and an ansatz analogous to that of
(\ref{sol}) leads to the following wave functions
\be
\phi (\vec{x} , t) = e^{-i \lambda \omega t} \; Y^{l}_{m}(\theta , \varphi)
(1 + \frac{\omega^2}{c^2} r^2)^{- \frac{\lambda}{2}} \; r^l
\phi^{\lambda}_{l}(r) \, ,
\ee
where $Y^{l}_{m}(\theta , \varphi)$ are the spherical harmonics and the
functions $\phi^{\lambda}_{l}(r)$ are required to verify the equation
\bea
\lefteqn{ \left\{ (1 + \frac{\omega^2}{c^2} r^2) \frac{d^2}{dr^2}
- 2 \frac{\omega^2}{c^2} r (\lambda - l - 2)
\frac{d}{dr} + 2 \frac{l+1}{r} \frac{d}{dr} + \right. } \nonumber \\
                             & &
\left. + \frac{\omega^2}{c^2} \left( \lambda (\lambda - 2 l - 3) + l(l + 3)
- N^2 + 12 \xi \right) \right\} \phi^{\lambda}_{l}(r) = 0 \, .
\eea
In terms of the variable $\rho = - \frac{\omega^2}{c^2} r^2$, the above
equation turns out to be a hypergeometric differential equation
\bea
\label{hiper3}
\lefteqn{ \left\{ (1 - \rho) \rho \frac{d^2}{d\rho^2} + \left( l +
\frac{3}{2} - (l + \frac{5}{2} - \lambda) \rho \right)
\frac{d}{d\rho} - \right. } \nonumber \\
                               & &
\left. - \frac{1}{4} \left( \lambda(\lambda - 2 l - 3) + l(l + 3) - N^2 +
12 \xi \right) \right\} \phi^{\lambda}_{l}(\rho) = 0 \, .
\eea
The regularity condition at the origin and the square integrability of the
wave functions yield to the following energy spectrum
\be
\label{espectro}
E = \left( \frac{3}{2} + 2 n + l + \frac{1}{2} \sqrt{9 + 4 \frac
{m^2 c^4}{\hbar^2 \omega^2} - 48 \xi} \right) \, \hbar \omega \, ,
\ee
where $n , l = 0 , 1 , 2 , 3 , \ldots$. We observe again that the energy
spectrum coincides, up to the zero-point energy, with that of the
non-relativistic limit. In the non-relativistic limit $c \rightarrow \infty$
the spectrum behaves as
\be
E \; \stackrel{c \rightarrow \infty}{\longrightarrow} \; E^{NR} + m c^2 \, ,
\ee
where $E^{NR} = (\frac{3}{2} + 2 n + l) \, \hbar \omega$ is the ordinary
energy spectrum of three-dimensional non-relativistic oscillator. \\

	For the values (\ref{espectro}) the regular hypergeometric functions
solving (\ref{hiper3}) are
\be
\label{F}
\phi^{\lambda}_{nl}(r) = \,_{2}F_{1}(-n , n + l + \frac{3}{2} - \lambda , l +
\frac{3}{2} ; - \frac{\omega^2}{c^2} r^2 ) \, ,
\ee
where $\lambda = \frac{E}{\hbar \omega}$.Rewritting the dimensionless
parameter $\frac{\omega^2}{c^2} r^2$ as $\frac{m \omega}{\hbar N} r^2$  it
is easy to check that, in the limit $c \rightarrow \infty$, the functions
(\ref{F}) become the confluent hypergeometric ones appearing in the
non-relativistic wave functions
\be
\lim_{c \rightarrow \infty} \,_{2}F_{1}(-n , n + l + \frac{3}{2} - \lambda , l
+
\frac{3}{2} ; - \frac{m \omega}{\hbar} \frac{r^2}{N}) = \,_{1}F_{1}(-n , l +
\frac{3}{2} ; \frac{m \omega}{\hbar} r^2) \, .
\ee

	Taking into account the relation of Jacobi polynomials with the
hypergeome\-tric functions \cite{Abram}
\be
_{2}F_{1}(-n , n + \alpha + \beta + 1 , \alpha + 1 ; z) = \frac{n!}{(\alpha +
1)_{n}}
P_{n}^{(\alpha , \beta)}(1 - 2z) \, ,
\ee
a orthonormal basis for the Hilbert space is given by
\be
\Psi^{\gamma}_{nlm}(\vec{x} , t) = C_{nlm}^{\gamma} \; e^{-i \frac{E}{\hbar} t}
\; Y^{l}_{m}(\theta , \varphi) \, (1 + \frac{\omega^2}{c^2} r^2)^{- \frac
{\lambda + 1}{2}} \; r^l P^{[l + \frac{1}{2} , - \lambda]}_{n}(1 + 2 \frac
{\omega^2}{c^2} r^2) \, .
\ee
where $C_{nlm}^{\gamma}$ are normalization constants.
Observe that the polynomials $P_{n}^{[l + \frac{1}{2} , - \lambda]}$ have
the appropiate $c \rightarrow \infty$ limit
\be
\lim_{c \rightarrow \infty} P_{n}^{[l + \frac{1}{2} , - \lambda]}(1 + 2
\frac{m \omega}{\hbar} \frac{r^2}{N}) = L_{n}^{(l + \frac{1}{2})}(\frac
{m \omega}{\hbar} r^2) \, ,
\ee
where $L_{n}^{(l + \frac{1}{2})}$ are the generalized Laguerre polynomials.
To illustrate our relativistic generalization of them we give the first
few polynomials
\footnotesize
\bea
P_{0}^{[l + \frac{1}{2} , -(l + \frac{3}{2} + \gamma)]}(1 + 2 \frac{\zeta^2}
{N}) & = & 1 \\
P_{1}^{[l + \frac{1}{2} , -(l + \frac{5}{2} + \gamma)]}(1 + 2 \frac{\zeta^2}
{N}) & = & (l + \frac{3}{2}) - 2 \frac{\gamma}{N} \zeta^2 \\
P_{2}^{[l + \frac{1}{2} , -(l + \frac{7}{2} + \gamma)]}(1 + 2 \frac{\zeta^2}
{N}) & = & \frac{1}{2} (l + \frac{3}{2}) \, (l +\frac{5}{2}) - \frac{1}{2}
\frac{\gamma (l + \frac{5}{2})}{N} \zeta^2 +
\frac{1}{8} \frac{\gamma (\gamma + 1)}{N^2} \zeta^4 \, ,
\eea
\normalsize
where now $\gamma = \frac{1}{2} \, \sqrt{9 + 4 N^2 - 48 \xi}$. \\
\\
\\

\section{Conclussions and final comments}

\hspace{5mm} In this paper we have shown that a particular anti-de Sitter
metric
can be used to simulate, in a natural way, a relativistic harmonic oscillator.
The system is exactly solvable and leads to radial energy eigen-functions
composed of a weight-function
\be
\label{peso}
\left( 1 + \frac{\omega^2}{c^2} r^2 \right)^{\left( -\frac{1}{4} \sqrt{9 +
4 \frac{m^2 c^4}{\hbar^2 \omega^2} - 48 \xi} - \frac{1}{2} (\frac{3}{2} +
2 n + l) \right)} \, ,
\ee
reducing to the Gaussian one $e^{- \frac{1}{2} \frac{m \omega}{\hbar} r^2}$
in the limit $c \rightarrow \infty$, and a polynomial
\be
r^l \, P_{n}^{[l + \frac{1}{2}, - \lambda]}(1 + 2 \frac{\omega^2}{c^2}
r^2) \, ,
\ee
going to its non-relativistic counterpart. \\

	We observe from the expression (\ref{peso}) that the probability
density for the re\-la\-tivistic oscillator is less confined in the classical
region than the corresponding one of the non-relativistic oscillator. It
penetrates more appreciably in the classically  forbiden region. This can be
understood in terms of the behaviour of the null geodesics in AdS. They go to
infinity in a finite lapse of coordinate time. So that, in the limit
$N = 0$, the geodesics are not confined in a finite region of space and this
fact is partialy reflected by the asymptotic behaviour of the
wave functions. Despite of this, the spacing of the energy levels is identical
to the non-relativistic one. However for the ground state energy we have
\be
E_0 = \hbar \omega \left( \frac{3}{2} + \frac{1}{2} \, \sqrt{9 + 4 N^2 -
48 \xi} \right) \, ,
\ee
representing some sort of mixing between the non-relativistic zero-point
energy $\frac{3}{2} \hbar \omega$ and the relativistic rest mass energy.
The mixing is just parametrized by the curvature factor $\xi$. \\

	Another point which merits some comments is the question of how to
extend our approach to spinning particles. When dealing with spin $\frac{1}{2}$
particles one could construct the corresponding wave equation by means of
Dirac equation in the AdS background, i.e.
\be
\label{dirac}
\left( i \gamma^{\mu} (\partial_{\mu} - \Gamma_{\mu}) - \frac{mc}{\hbar}
\right) \, \Psi = 0 \, ,
\ee
where $\gamma^{\mu} = e^{\mu}_{\; \alpha} \, \gamma^{\alpha}$ are the Dirac
matrices in AdS space ($e^{\mu}_{\; \alpha}$ are the vierbeins) and
$\Gamma_{\mu}$ is the spin connection. Using the identity
\be
R_{\mu \nu \sigma \rho} \; \gamma^{\mu} \gamma^{\nu} \gamma^{\sigma}
\gamma^{\rho} \, = - 2 R \, ,
\ee
where $R_{\mu \nu \sigma \rho}$ is the Riemann curvature tensor, it is not
dificult to see that the Dirac equation (\ref{dirac}) implies a Klein-Gordon
equation with $\xi = \frac{1}{4}$
\be
\label{kg2}
\left( \square + \left (\frac{mc}{\hbar} \right)^2 + \frac{1}{4} R \right)
\, \Psi = 0 \, .
\ee
The term $\frac{1}{4} R$ plays the role of the standard spin-dependent term
$\frac{q}{2} F_{\mu \nu} \sigma^{\mu \nu}$ which appears when coupling the
Dirac field with a electromagnetic potential. Note that this term is now
diagonal in the spin components and then does not yield to a spin-orbit
coupling. \\
\\
\\

\section*{Acknowlegements}

\hspace{5mm} J. Navarro-Salas would like to thank V. Aldaya and
J. Guerrero for valuable discusions. \\
\\
\\

\end{document}